\documentclass[prl,twocolumn,showpacs,superscriptaddress,reprint]{revtex4}%
\usepackage{graphicx}
\usepackage{latexsym}
\usepackage{amsmath}
\usepackage{amssymb}
\usepackage{amsfonts}
\usepackage{color}%
\providecommand{\U}[1]{\protect\rule{.1in}{.1in}}

\begin{document}
\title{Chirality-controlled spontaneous currents in spin-orbit coupled
superconducting \ rings}
\author{J. W. A. Robinson}\email{jjr33@cam.ac.uk}
\affiliation{Department of Materials Science and Metallurgy,
University of Cambridge, CB3 0FS Cambridge,United Kingdom}
\author{A. V. Samokhvalov}\email{samokh@ipmras.ru}
\affiliation{Institute for Physics of Microstructures, Russian
Academy of Sciences, 603950 Nizhny Novgorod, GSP-105, Russia}
\affiliation{Lobachevsky State University of Nizhni Novgorod, 23
Prospekt Gagarina, 603950 Nizhni Novgorod, Russia} \affiliation{}
\author{A. I. Buzdin}\email{alexandre.bouzdine@u-bordeaux.fr}
\affiliation{University Bordeaux, LOMA UMR-CNRS 5798, F-33405
Talence Cedex, France} \affiliation{Department of Materials
Science and Metallurgy, University of Cambridge, CB3 0FS
Cambridge,United Kingdom}
\date{\today}

\begin{abstract}

\end{abstract}
\begin{abstract}
At a superconductor interface with a ferromagnetic insulator (FI), the FI acts
to induce a local exchange field within the S layer, which in the presence of
spin-orbit interaction promotes a phase- modulated superconducting state. Here
we demonstrate that within a thin superconducting loop that is partially
proximitized by a FI, spontaneous currents form with a
magnetization-orientation-dependent chirality with sizable shifts in
Little-Parks oscillations. Furthermore, the critical temperature of the loop
is also magnetization-orientation-dependent and conversely, the
superconducting transition itself may influence the magnetization direction.
More generally, the superconducting region above the FI may serve as a
``phase battery" and so offer a new device concept for
superconducting spintronics.

\end{abstract}

\maketitle

\bigskip

The interaction of interface superconductivity with materials exhibiting
strong spin-orbit coupling \cite{Mineev_Review,Agterberg_review} and magnetic
exchange fields \cite{Robinson1} offers enormous potential for the discovery
and control of new physical phenomena. For example, a homogeneous magnetic
exchange field acting at a superconductor/ferromagnet (S/F) interface induces
oscillations in the superconducting order parameter in S/F/S Josephson
junctions \cite{SFReviews,SFSjuntions1,SFSjuntions2,SFSjuntions3} whilst a
non-uniform exchange field in such junctions can lead to electron pair
conversion from spin-singlet to spin-triplet \cite{Robinson2,Birge1} and a
dependence of the superconducting critical temperature $T_{c}$ on
magnetization alignment in S/F/F multilayers \cite{SFmultipairconversion}. In
the absence of inversion symmetry, Rashba spin-orbit interaction (SO) in
combination with a magnetic exchange field or Zeeman field offers additional
physics with an unusual linear over the gradient of the superconducting order
parameter $\Psi$ terms in the Ginzburg-Landau (GL) free energy $\sim\left(
\nabla\Psi\right)  \Psi^{\ast}$. Here we call such coupling the
\textquotedblleft exchange spin-orbit coupling\textquotedblright\ or
\textquotedblleft EXSO\textquotedblright. The EXSO effect induces different
types of Larkin-Ovchinnikov-Fulde-Ferrell (LOFF)-like helical phases with a
non-zero Cooper pair momentum $\vec{p}$ in the ground state
\cite{Barzykin,Samokhin_SOC, Kaur_SOC, Edelstein_PRL, Dimitrova}, which play
an important role in Majorana physics \cite{Alicea} and lead to the formation
of $\varphi_{0\text{ }}$-Josephson junction with a spontaneous phase
difference of $\varphi_{0\text{ }}$ in the ground state
\cite{Buzdin_Phi,Krive,Reynoso}. Recently, such $\varphi_{0\text{ }}%
$-Josephson have\ been realized experimentally \cite{Kouwenhoven}.

We note that in uniform systems the helical phases do not carry a current;
however, such current-carrying states may appear in non-uniform EXSO systems
such as close to a magnetic island on a thin film superconductor
\cite{Balatskii} or near an S/F interface within a distance of the London
penetration depth of the interface \cite{Mironov}.

\begin{figure}[ptb]
\includegraphics[width=0.3\textwidth]{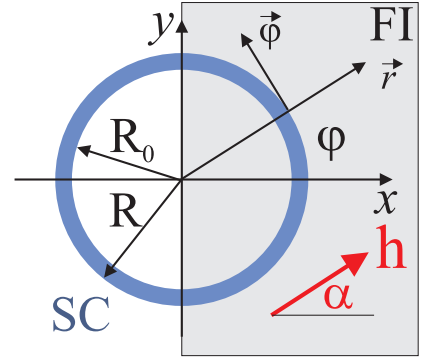}\caption{(Color online) A thin-film superconducting loop proximity-coupled to a
ferromagnetic insulator (FI) occupying the half plane $x\geq0$. $R$ ($R_{0}$)
is the external (internal) radius of the loop ($R-R_{0}\leq\xi$),
$\mathbf{h}=h\mathbf{x}_{0}$ is the exchange field, and $(r,\,\varphi)$ is the
polar coordinate system.}%
\label{Fig:Struct}%
\end{figure}

In this \textit{letter} we demonstrate that the EXSO effect leads to
spontaneous currents in a closed superconducting loop in which the
superconductor is partially-coupled to a ferromagnetic insulator (FI) as shown
in Fig.~\ref{Fig:Struct}. We further demonstrate that the chirality of the
current is controllable through the magnetization alignment of the FI. The
center of the ring of radius $R$ is at $x=y=0$. In this geometry, we expect a
non-trivial interplay between the Little-Parks effect and helical current
carrying states. Even in zero external magnetic field, spontaneous currents
are generated in the ground state. The study of Little-Parks oscillations
is a powerful tool to probe subtle effects and was recently applied to
investigate the superconducting symmetry in Sr$_{2}$Ru0$_{4}$ microrings
\cite{Yasui-Science11}.
%

To describe the superconducting loop we apply the general Ginzburg-Landau
(GL) approach which is relevant at a temperature $T$ close to the critical
transition $T_{c}$ of the loop. In the presence of the EXSO, the density
$f(\mathbf{r})$ of the GL free energy $F=\int f(\mathbf{r})d^{3}\mathbf{r}$
reads \cite{Samokhin_SOC, Kaur_SOC}
\begin{align}
f(\mathbf{r})  &  =a|\psi|^{2}+\gamma|\hat{\mathbf{D}}\,\psi|^{2}+\frac{b}%
{2}|\psi|^{4}\label{Initial functional}\\
&  +\left[  \vec{\mathbf{n}}\times\vec{\mathbf{h}}\right]  \cdot\left[
\psi^{\ast}\varepsilon_{0}(\mathbf{r})\,\hat{\mathbf{D}}\,\psi+\mathrm{c.c.}%
\right]  \,.\nonumber
\end{align}
Here $a=-\alpha(T_{c}-T)$, $\alpha$, $b$ and $\gamma>0$ are the standard GL
coefficients, $\psi$ is the superconducting order parameter with,
$\hat{\mathbf{D}}=-i\hbar\nabla+(2e/c)\mathbf{A}$ is the gauge-invariant
momentum operator ($e>0$), $\vec{n}$ is the unit vector in the direction along
which the inversion symmetry is broken (pointing in the $z$ direction
perpendicular to S/F interface in our case), $\vec{\mathbf{h}}$ is the
exchange field, and $\varepsilon_{0}(\mathbf{r})$ is the EXSO constant. We
assume that the exchange field $\vec{\mathbf{h}}$ in superconductor is
generated by the surface field of the ferromagnetic insulator and as Rashba SO
interaction they are localized near the S/F interface. We consider the case of
the thin ring (thickness and width smaller than superconducting coherence
length $\xi$) where the SO and Zeeman fields are effectively averaged over the
cross section of the ring. Taking into account that in our case both the SO
interaction and exchange field are present at the distance of the order of
interatomic distance $a$ from the S/F interface and then they are averaged
over the ring thickness, we have ${\varepsilon}$ $\sim\left(  a/d\right)
{\varepsilon}_{b}$, where $d$ is the thickness of the superconductor and
${\varepsilon}_{b}$ $\sim\gamma\,v_{so}/v_{F}^{2}$ is the estimation of the
EXSO constant for the bulk superconductor, see for example \cite{Dimitrova}.
Here $v_{so}$ is the characteristic velocity entering in SO Rashba interaction
and $v_{F}$ is the Fermi velocity.

For a $1D$ ring of radius $R$ it is convenient to write the GL free energy
functional (\ref{Initial functional}) using the polar coordinates
($r,\,\varphi$), and take into account that the tangential $\varphi-$component
$A$ of the vector-potential $\mathbf{A}=(0,\,A,\,0)$ is directly related with
the magnetic flux $\Phi=2\pi RA$ enclosed by the superconducting loop. Then
the GL free energy then reads
\begin{align}
F  &  =\sigma R\int\limits_{0}^{2\pi}d\varphi\left\{  a|\psi|^{2}+\gamma
\frac{\hbar^{2}}{R^{2}}\bigg\vert\left(  -i\frac{\partial}{\partial\varphi
}+\phi\right)  \psi\bigg\vert^{2}\right. \label{functional}\\
&  +\left.  \frac{b}{2}|\psi|^{4}+\frac{\hbar}{R}\,\varepsilon(\varphi)\left[
\psi^{\ast}\left(  -i\frac{\partial}{\partial\varphi}+\phi\right)  {\psi
}+\mathrm{c.c.}\right]  \right\}  \,.\nonumber
\end{align}
where $\sigma$ is the cross section of the ring, $\phi=\Phi/\Phi_{0}$ is the
enclosed magnetic flux in the units of the flux quantum $\Phi_{0}=\pi\hbar
c/e$ and
\[
\varepsilon(\varphi)=\varepsilon_{0}\left[  \vec{\mathbf{n}}\times
\vec{\mathbf{h}}\right]  _{\varphi}=h\varepsilon_{0}\cos(\varphi-\alpha)
\]
is a projection on the ring of the Rashba SOC for $|\varphi|\leq\pi/2$ and
${\varepsilon(\varphi)=0}$ otherwise.

To calculate $T_{c}$ we are concerned with the linear equation for the
superconducting order parameter, which for the functional (\ref{functional})
can be written as
\begin{equation}
\left(  \frac{\partial}{\partial\varphi}+i\phi\right)  ^{2}\psi+2i\lambda
(\varphi)\left(  \frac{\partial\psi}{\partial\varphi}+i\phi\,\psi\right)
+i\frac{\partial\lambda}{\partial\varphi}\,\psi=\tau\psi\,, \label{GL}%
\end{equation}
where %
$\lambda(\varphi)=\lambda_{so}\cos(\varphi-\alpha)$, %
$\lambda_{so}=\varepsilon_{0}hR/\gamma\,\hbar$ %
is the EXSO parameter for $|\varphi|\leq\pi/2$ and
${\lambda(\varphi)=0}$ otherwise. The eigenvalue $\tau$ determines
the shift of the transition temperature
\begin{equation}
T_{c}=T_{c0}\left(  1+\frac{\xi_{0}^{2}}{R^{2}}\,\tau\right)
\label{TransTemper}%
\end{equation}
with respect to its bulk value $T_{c0}$. Here $\xi_{0}^{2}=\gamma\hbar
^{2}/\alpha T_{c0}$ is the GL coherence length. Note that averaged exchange
field $h_{av}\sim h\,a/d$ and to preserve the superconductivity we need
$h_{av}\lesssim T_{c}$. However at $v_{so}/v_{F}\sim\mathrm{0.1}$ we may have
for $R/\xi_{0}\sim\mathrm{10\div100}$ a pretty large spin-orbit parameter
$\lambda_{so}\sim\mathrm{1\div10}$ .

The current density corresponding to the functional (\ref{functional}) is
determined by the relation
\begin{equation}
j=-\frac{{\partial f}}{\partial A}=\frac{2e\gamma\,\hbar}{R}\left[  i{\psi
}\frac{{\partial\psi^{\ast}}}{\partial\varphi}-i{\psi^{\ast}}\frac
{{\partial\psi}}{\partial\varphi}+2\lambda{\psi^{\ast}{\psi}}\right]
\label{CurrDens}%
\end{equation}
and we may verify that the Eq.~(\ref{GL}) guarantees $\operatorname{div}%
\mathbf{j}=0$.

To explain our main results, we start from a qualitative discussion of the SOC
effect on the Little-Parks oscillations. By introducing an effective flux
$\tilde{\phi}(\varphi)=\phi+\lambda(\varphi)$, we write the equation
(\ref{GL}) in following form
\[
\left(  \partial_{\varphi}+i\tilde{\phi}\right)  ^{2}\psi=\tilde{\tau}\psi\,,
\]
where $\tilde{\tau}(\varphi)=\tau-\lambda^{2}(\varphi)$. Then the averaged
over the ring value
\[
\langle\,\tilde{\phi}\,\rangle=\frac{1}{2\pi}\int\limits_{0}^{2\pi}%
d\varphi\,\tilde{\phi}(\varphi)=\phi+(\lambda_{so}/\pi)\cos\alpha\,,
\]
describes the additional shift
\begin{eqnarray}\label{FluxShift}%
    \Delta \phi = - ( \lambda_{so} / \pi ) \cos\alpha
\end{eqnarray}
of the $T_{c}$ maximum of the Little-Parks oscillations.

To provide the full solution of (\ref{GL}) we use the Fourier series expansion
for the order parameter
\begin{equation}
\psi=\sum\limits_{n}\psi_{n}\exp(-in\varphi) \label{FourPsi}%
\end{equation}
and EXSO coupling constant
\begin{equation}
\lambda(\varphi)=\sum\limits_{n}\lambda_{n}\exp(-in\varphi)\,,
\label{FourLamb}%
\end{equation}
where
\begin{eqnarray}
\lambda_{n}  &=& \lambda_{n}^{r}+i\,\lambda_{n}^{i}\,
    = \frac{\lambda_{so}\cos\alpha}{2\pi}\int\limits_{-\pi/2}^{\pi/2}%
        d\varphi\,\cos\varphi\,\cos(n\varphi) \nonumber \\
            &+& i\frac{\lambda_{so}\sin\alpha}{2\pi}%
                \int\limits_{-\pi/2}^{\pi/2}d\varphi\,\sin\varphi\,\sin(n\varphi) \label{FourLamb1}
\end{eqnarray}
Then we readily obtain from the Eq.~(\ref{GL}) the systems of
coupled algebraic equations for the harmonics ${\psi}_{n}$
\begin{equation}
-\left(  n-\phi\right)  ^{2}\psi_{n}+\sum\limits_{m}\lambda_{n-m}\left(
n+m-2\phi\right)  \psi_{m}=\tau\,\psi_{n}\,. \label{SystemPsi}%
\end{equation}

Let us consider first the case of the flux smaller than the half quantum
$\phi<1/2$. In this situation zero harmonics $\psi_{0}$ will dominate (
$\psi_{n}\ll\psi_{0})$ if $\lambda_{so}$ is small:
\[
\left(  \tau+\phi^{2}+2\phi\lambda_{0}\right)  \psi_{0}=\underset{n\neq
0}{\sum}\lambda_{n}^{\ast}\,(n-2\phi)\,\psi_{n}\,.
\]
On the other hand for $\psi_{n}$ , we have
\[
\left[  \,\tau+(n-\phi)^{2}-2(n-\phi)\lambda_{0}\,\right]  \psi_{n}%
\approx\lambda_{n}(n-2\phi)\,\psi_{0}\,,
\]
which leads to the equation for $\tau$
\[
\left(  \tau+\phi^{2}+2\phi\lambda_{0}\right)  =\underset{n\neq0}{\sum}%
\frac{|\lambda_{n}|^{2}(n-2\phi)^{2}}{\tau+(n-\phi)^{2}-2(n-\phi)\lambda_{0}%
}\,.
\]
Since $\tau\approx-\phi^{2}-2\lambda_{0}\phi$ and than $\tau+(n-\phi
)^{2}-2\lambda_{0}(n-\phi)\approx n[n-2(\lambda_{0}+\phi)]\approx n(n-2\phi)$
and finally we have for the transition temperature
\[
\tau=-\phi^{2}-2\lambda_{0}\phi+\underset{n\neq0}{\sum}|\lambda_{n}%
|^{2}=-(\phi+\lambda_{0})^{2} + \lambda_{so}^{2} / 4,
\]
where we took into account that $\lambda_{-n}=\lambda_{n}^{\ast}$ and
\[
\underset{n}{\sum}|\lambda_{n}|^{2}=2|\lambda_{1}|^{2}+\underset{m}{\sum
}|\lambda_{2m}|^{2} = \lambda_{so}^{2} / 4 \,.
\]
In the absence of the external flux ($\phi=0$) we have the following
dependence of the critical temperature $\tau$ on the orientation of the
exchange field $\mathbf{h}$:
\begin{equation}
\tau=\lambda_{so}^{2}\left[  1-\left(  2\cos\alpha/\pi\right)  ^{2}\right]
/4\,. \label{tau-al}%
\end{equation}
In general case if the flux $\phi$ is large $\min\left\vert n-\phi\right\vert
$ occurs for some finite $n_{0}$ and for $\alpha=0$ we have similarly
\[
\tau=-(\phi+\lambda_{0}-n_{0})^{2} + \lambda_{so}^{2} / 4 \,.
\]

\begin{figure}[ptb]
\includegraphics[width=0.35\textwidth]{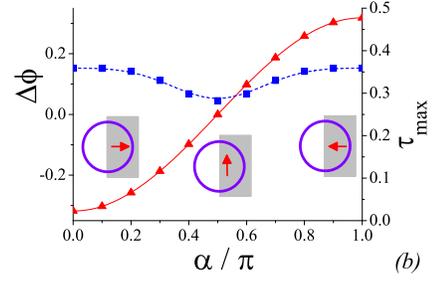}
\includegraphics[width=0.5\textwidth]{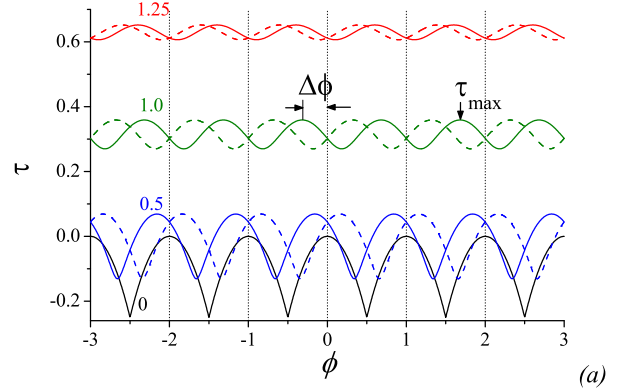}\caption{(Color online) ($a$)
Little-Parks oscillations for different values of EXSO parameter
$\lambda(\varphi)=\pm\lambda_{so}\cos(\varphi-\alpha)$ for $|\varphi|\leq
\pi/2$: $\alpha=0$ -- solid; $\alpha=\pi$ -- dashed. The numbers near the
curves denote the corresponding values of the spin-orbit constant
$\lambda_{so}$. ($b$) Dependence of the shifts of the Little-Parks
oscillations $\Delta\phi$ ({\color{red} $\blacktriangle$}) and the maximal
transition temperature $\tau_{max}$ ({\color{blue} $-\blacksquare-$}) (see the
panel ($a$) in confusion) on the angle $\alpha$ for $\lambda_{so}=1$. Solid
red line shows the dependence described by the Eq.~(\ref{FluxShift}).}%
\label{Fig:LP}%
\end{figure}
\begin{figure}[ptb]
\includegraphics[width=0.45\textwidth]{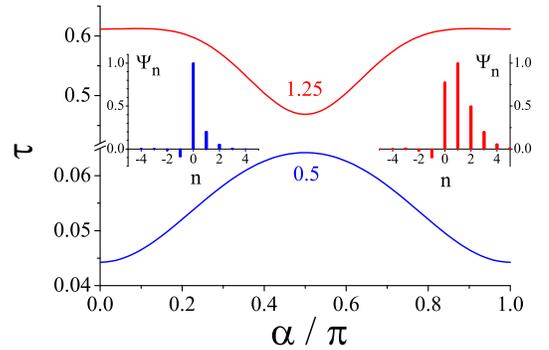}\caption{(Color online)
Dependence of the transition temperature $\tau$ on the angle $\alpha$ for
$\lambda_{so}=0.5$ (blue solid line) and $\lambda_{so}=1.25$ (red solid line)
obtained by the numerical solution of the eigenvalue problem (\ref{SystemPsi})
for $\phi=0$. The inserts show the spectra of orbital harmonics for $\alpha=0$
and different values of the spin-orbit constant: left -- $\lambda_{so}=0.5$;
right -- $\lambda_{so}=1.25$. }%
\label{Fig:Spec}%
\end{figure}

The results of the numerical calculations for arbitrary values of
$\lambda_{so}$ are plotted in Fig.~\ref{Fig:LP}$a$. We see that the strong
EXSO interaction shifts the Little-Parks oscillations with a dependence on the
orientation of $\mathbf{h}$. One can also observe the stimulation of the
superconductivity by spin-orbital interaction: in Fig.~\ref{Fig:LP}$b$ we
compare the dependence of the shifts $\Delta\phi$ of the $T_{c}$ maximum of
the Little-Parks oscillations obtained analytically, i.e., using
Eq.~(\ref{FluxShift}), and numerically. One can see an excellent agreement
between the qualitative description (\ref{FluxShift}) and the exact solution
of the eigenvalue problem (\ref{SystemPsi}). Note that the numerical
calculation demonstrates a weak dependence of the maximal $T_{c}$ on direction
of $\mathbf{h}$. In Figure~\ref{Fig:Spec} we show the dependence of the
transition temperature $\tau$ on the $\mathbf{h}$ direction $\alpha$ in the
absence of external flux ($\phi=0$) for two values of spin-orbital constant
$\lambda_{so}$. The insets in Figure~\ref{Fig:Spec} show spectra of orbital
harmonics for the selected values of $\lambda_{so}$ at $\alpha= 0$. For small
values of the EXSO constant the zero harmonic prevails (left insert in
Fig.~\ref{Fig:Spec}), and $\tau(\alpha)$ dependence is well described by the
expression (\ref{tau-al}). The spin-orbital interaction results in generation
of harmonics with nonzero orbital momenta. As a result the spectrum of orbital
harmonics spreads with an increase of the spin-orbital interaction, and a
dominant harmonic has a nonzero orbital momenta (right insert in
Fig.~\ref{Fig:Spec}).

\begin{figure}[ptb]
\includegraphics[width=0.45\textwidth]{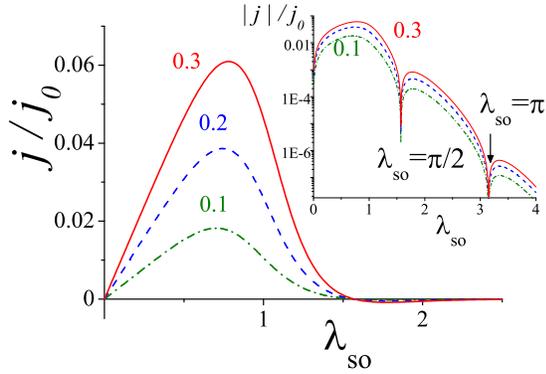} \caption{(Color online)
The dependence of the current density $j$ on spin-orbital constant
$\lambda_{SO}$ for different temperatures $T<T_{c}$ ( $j_{0}=4e\,(\hbar
/R)^{3}\gamma^{2}/b$ ). The numbers by the curves correspond to the
dimensionless shift $\Delta\tau=(R/\xi_{0})^{2}(T_{c}-T)/T_{c}$ in $T$ with
respect to the transition (\ref{TransTemper}). }%
\label{Fig:CD}%
\end{figure}

We now calculate the current for the case of spin-orbit interaction in the
absence of the applied field with $\phi=0$ and the exchange field oriented
along the $x$-axis ($\alpha=0$). We calculate the current at ${\varphi=}{\pi}%
$, where the expression for the current density has the standard form
(\ref{CurrDens}) \ with $\lambda=0$ and
\begin{equation}
j=-4e\frac{\gamma\,\hbar}{R}\underset{n\,m}{\sum}n\,(-1)^{n-m}\mathrm{Re}%
[\psi_{n}\psi_{m}^{\ast}]\approx4e\frac{\gamma\,\hbar\lambda_{so}}{\pi
R}\,\psi_{0}^{2} \,. \label{CurrDens-pi}%
\end{equation}
Figure~\ref{Fig:CD} shows the dependence the current density $j$
(\ref{CurrDens-pi}) on the spin-orbital constant $\lambda_{SO}$ for different
temperatures $T<T_{c}$. One can see that even in the absence of an external
magnetic field the superconducting transition occurs in the current-carrying
state with nonzero orbital momenta. The sign and magnitude of the spontaneous
supercurrent due to spin-orbital interaction is determined by the EXSO
constant with a maximum at $\lambda_{so} \sim1$.
%

\begin{figure}[t]
\includegraphics[width=0.2\textwidth]{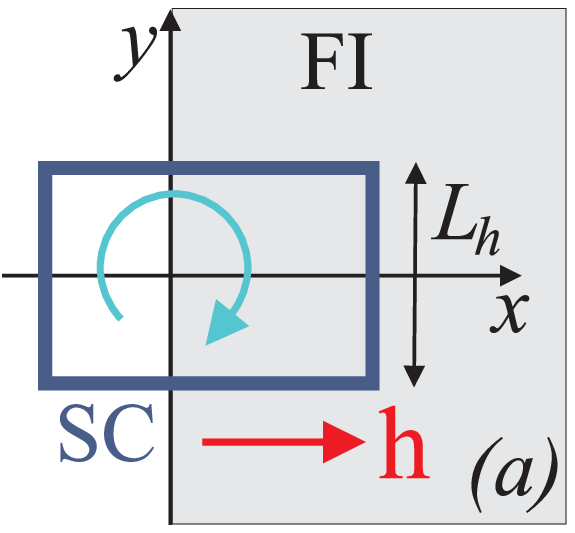}\hspace{1 cm}
\includegraphics[width=0.2\textwidth]{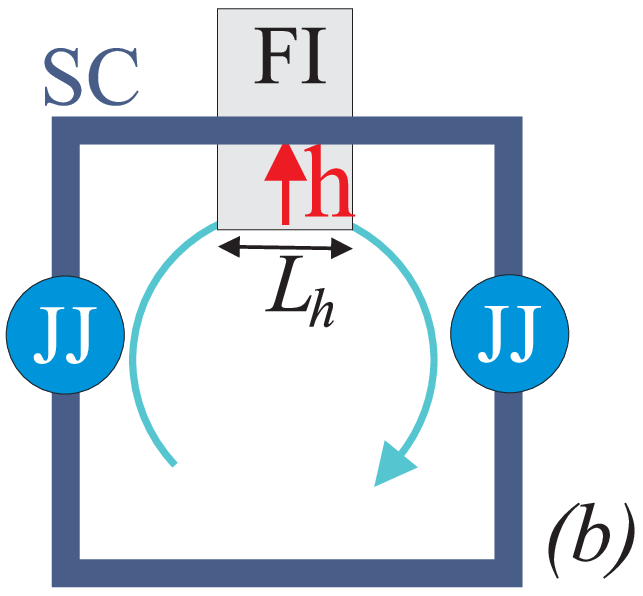} \caption{(Color online)
($a$) Schematic representation of a thin superconducting loop with the EXSO
interaction. ($b$) DC SQUID with the EXSO "phase battery".}%
\label{Fig:PhBat}%
\end{figure}

We have considered above the case of a thin superconducting ring; however, the
effect of the phase accumulation due to EXSO coupling (describing by the
constant $\varepsilon\sim h\varepsilon_{0}$ ) is very general and can play the
role of a superconducting phase battery. Indeed, let us consider the
superconducting loop at temperatures below $\ T_{c}\ \ $where$\ $modulus
$\left\vert {\psi}\right\vert $ of the order parameter is constant. In the
case of the infinite wire in the presence of the EXSO coupling the phase
modulated solution is realized ${\psi=}\left\vert {\psi}\right\vert
\exp(i\varphi)$ with $\varphi=q_{0}l\left\vert {\psi}\right\vert \ $and
$q_{0}=h\varepsilon_{0}/\gamma\hbar$ ($l$ - coordinate along the wire). This
solution, however, corresponds to the absence of superconducting current
\begin{align}
j  &  = 2ie\hbar\gamma\,(\psi\frac{\partial\psi^{\ast}}{\partial l}-\psi
^{\ast} \frac{\partial\psi}{\partial l}) - 4 e h \varepsilon_{0}|\psi
|^{2}\nonumber\\
&  = 4 e \hbar\gamma|\psi|^{2}(\frac{\partial\varphi}{\partial l}-q_{0}) =
0\,.\nonumber
\end{align}
Boundary condition at the interface between modulated ($\varepsilon\neq0)$ and
usual superconducting phase ($\varepsilon=0)$ corresponds simply to the
continuity of the current. Let us consider as a generic example of the
superconducting wire loop consisting of the region of the length $L_{h}$ with
exchange field and $L_{0}$ without exchange field (see Fig.~\ref{Fig:PhBat}%
$a$). The solution in the first region $\left\vert {\psi}\right\vert
\exp(iq_{1}l)$ and the corresponding current $j_{1}=4e\hbar\gamma|\psi
|^{2}(q_{1}-q_{0})$. The solution in the second region $|\psi|\exp(iq_{2}l)$
and the corresponding current $j_{2}=4e\hbar\gamma|\psi|^{2}q_{2}$. From the
continuity of the current we have $q_{1}-q_{0}=q_{2}$. The total phase
accumulation over the loop will be $\Delta\varphi=q_{1}L_{h}+q_{2}L_{0}$ and
should be equal to $2\pi n$. From this condition we find $q_{2}=(2\pi
n-q_{0}L_{h})/(L_{h}+L_{0})$, what means the spontaneous generation of the
current ( except the special cases when the phase accumulation in active
$L_{h}$ region is integer of $2\pi$, i.e. $q_{0}L_{h}=2\pi n$ ). The actual
choice of $n$ is dictated by the minimization of the current $j_{1}=j_{2}$ and
so the associated magnetic energy, i.e. $\min\left\vert 2\pi n-q_{0}%
L_{h}\right\vert $. Therefore we may conclude that the part of the wire with
EXSO coupling serves as a "phase battery".
%

In conclusion, we have demonstrated that the spin-orbit interaction in
combination with a magnetic exchange field (EXSO) may induce non-dissipative
currents in superconducting loops with a chirality that is controllable
through the magnetization alignment. Superconducting elements with EXSO may
generate spontaneous phase shifts governed by magnetic moment orientation and
by introducing the EXSO wire $L_{h}$ in a DC SQUID (see Fig.~\ref{Fig:PhBat}%
$b$) we can mimic the role of external flux $2\pi\Phi/\Phi_{0}=q_{0}L_{h}$
which opens the way to creating "quiet" qubits \cite{Ioffe-QuQUBIT}, similar
to that which has been recently fabricated on the basis of the $\pi-$junction
\cite{pi-QuQUBIT}. Suitable FI materials for the proposed experiments here
include EuS \cite{FIMaterials1} and GdN \cite{FIMaterials2} and potentially
oxides such as praseodymium calcium manganese oxide \cite{FIMaterials3}.
Finally, we note that the EXSO interaction between magnetic and
superconducting subsystems may, in principle, generate an inverse effect in
which the superconducting transition provokes reorientation of the magnetic moment.

\bigskip

J.W.A.R. acknowledges funding through EPSRC-JSPS International Network and Programme Grants (No. EP/P026311/1 and No. EP/N017242/1). J.W.A R. also acknowledges funding from the
Royal Society through a University Research Fellowship and with A.
I. B., funding from the Leverhulme Trust and EU Network COST
CA16218 (NANOCOHYBRI). This work was supported by Russian
Foundation for Basic Research under Grants No. 17-52-12044 NNIO
and No. 18-02-00390 ( A. V. S.), the French ANR project
SUPERTRONICS and OPTOFLUXONICS (A. I. B.),


\end{document}